\documentclass{article}
\usepackage{amsmath, amssymb, amsfonts}
\title{ On a star with expanding isotropic fluid } 
\author{Hristu Culetu \footnote{electronic address: hculetu@yahoo.com}\\ Constanta, Romania}

\begin{document}
\numberwithin{equation}{section}
\pagenumbering{arabic}
\maketitle
\newcommand{\fv}{\boldsymbol{f}}
\newcommand{\tv}{\boldsymbol{t}}
\newcommand{\gv}{\boldsymbol{g}}
\newcommand{\OV}{\boldsymbol{O}}
\newcommand{\wv}{\boldsymbol{w}}
\newcommand{\WV}{\boldsymbol{W}}
\newcommand{\NV}{\boldsymbol{N}}
\newcommand{\hv}{\boldsymbol{h}}
\newcommand{\yv}{\boldsymbol{y}}
\newcommand{\RE}{\textrm{Re}}
\newcommand{\IM}{\textrm{Im}}
\newcommand{\rot}{\textrm{rot}}
\newcommand{\dv}{\boldsymbol{d}}
\newcommand{\grad}{\textrm{grad}}
\newcommand{\Tr}{\textrm{Tr}}
\newcommand{\ua}{\uparrow}
\newcommand{\da}{\downarrow}
\newcommand{\ct}{\textrm{const}}
\newcommand{\xv}{\boldsymbol{x}}
\newcommand{\mv}{\boldsymbol{m}}
\newcommand{\rv}{\boldsymbol{r}}
\newcommand{\kv}{\boldsymbol{k}}
\newcommand{\VE}{\boldsymbol{V}}
\newcommand{\sv}{\boldsymbol{s}}
\newcommand{\RV}{\boldsymbol{R}}
\newcommand{\pv}{\boldsymbol{p}}
\newcommand{\PV}{\boldsymbol{P}}
\newcommand{\EV}{\boldsymbol{E}}
\newcommand{\DV}{\boldsymbol{D}}
\newcommand{\BV}{\boldsymbol{B}}
\newcommand{\HV}{\boldsymbol{H}}
\newcommand{\MV}{\boldsymbol{M}}
\newcommand{\be}{\begin{equation}}
\newcommand{\ee}{\end{equation}}
\newcommand{\ba}{\begin{eqnarray}}
\newcommand{\ea}{\end{eqnarray}}
\newcommand{\bq}{\begin{eqnarray*}}
\newcommand{\eq}{\end{eqnarray*}}
\newcommand{\pa}{\partial}
\newcommand{\f}{\frac}
\newcommand{\FV}{\boldsymbol{F}}
\newcommand{\ve}{\boldsymbol{v}}
\newcommand{\AV}{\boldsymbol{A}}
\newcommand{\jv}{\boldsymbol{j}}
\newcommand{\LV}{\boldsymbol{L}}
\newcommand{\SV}{\boldsymbol{S}}
\newcommand{\av}{\boldsymbol{a}}
\newcommand{\qv}{\boldsymbol{q}}
\newcommand{\QV}{\boldsymbol{Q}}
\newcommand{\ev}{\boldsymbol{e}}
\newcommand{\uv}{\boldsymbol{u}}
\newcommand{\KV}{\boldsymbol{K}}
\newcommand{\ro}{\boldsymbol{\rho}}
\newcommand{\si}{\boldsymbol{\sigma}}
\newcommand{\thv}{\boldsymbol{\theta}}
\newcommand{\bv}{\boldsymbol{b}}
\newcommand{\JV}{\boldsymbol{J}}
\newcommand{\nv}{\boldsymbol{n}}
\newcommand{\lv}{\boldsymbol{l}}
\newcommand{\om}{\boldsymbol{\omega}}
\newcommand{\Om}{\boldsymbol{\Omega}}
\newcommand{\Piv}{\boldsymbol{\Pi}}
\newcommand{\UV}{\boldsymbol{U}}
\newcommand{\iv}{\boldsymbol{i}}
\newcommand{\nuv}{\boldsymbol{\nu}}
\newcommand{\muv}{\boldsymbol{\mu}}
\newcommand{\lm}{\boldsymbol{\lambda}}
\newcommand{\Lm}{\boldsymbol{\Lambda}}
\newcommand{\opsi}{\overline{\psi}}
\renewcommand{\tan}{\textrm{tg}}
\renewcommand{\cot}{\textrm{ctg}}
\renewcommand{\sinh}{\textrm{sh}}
\renewcommand{\cosh}{\textrm{ch}}
\renewcommand{\tanh}{\textrm{th}}
\renewcommand{\coth}{\textrm{cth}}

\begin{abstract}
The generalized Gullstrand-Painleve geometry is investigated for expanding matter. Compared to other studies, we take into account an anisotropic stress tensor as the source of curvature with an equation of state resembling the MIT bag model form. The spacetime becomes de Sitter for $t>>1/\sqrt{\Lambda},~ \Lambda$ being the equivalent cosmological constant. The energy density and pressures of the fluid are only time dependent but the scalar curvature is constant. The radial geodesics are computed.
 \end{abstract}
 
 \section{Introduction}
 The problem of gravitational collapse in general relativity leading eventually to black hole (BH) formation is of fundamental interest. Geometries with special symmetry are necessary to build gravitational collapse models, due to the complexity of Einstein's equations. Husain \cite{VH} gives exact non-static spherically symmetric solutions of the Einstein equations for a collapsing null fluid. He firstly determined the stress-energy tensor from the metric. Then the equation of state and the dominant energy condition
are imposed on its eigenvalues, so obtaining an equation for the metric function. A similar investigation on a regular dynamical metric is given in \cite{HC}, where the author proposed a metric with no singularities and an anisotropic stress tensor as the source of curvature.

 Vertogradov et al. \cite{VOS} found exact solutions to Einstein’s field equations that describe singularity-free spacetimes, through the gravitational collapse of baryonic matter characterized by a time- and radius-dependent coefficient of equation of state. The solutions have interesting physical properties but the pressure increases with radius and violates the dominant energy condition beyond some critical value. Vertogradov and Ovgun \cite{VO} investigated gravitational collapse scenarios for regular BH models in the transition from baryonic matter to quark matter. Certain strong interaction theories, particularly quark bag models, conjecture a phase of vacuum symmetry breaking within hadrons. This leads to significant differences between vacuum energy densities inside and outside hadrons, resulting in a vacuum-induced pressure on the quark confinement boundary (bag wall) \cite{FJ} (see also \cite{HC1}), which balances internal quark pressures and stabilizes the hadronic structure.

 Motivated by the previous achievements on the subject, we investigate the evolution of a high density star using a curved Gullstrand-Painleve geometry, due to a variable mass function $m(t,r)$, taking as source an imperfect fluid. The physical evolution of the system is viewed from the point of view of an observer comoving with the fluid.

The geometric units $c = G = 1$ are used, unless otherwise specified, where $c$ is the velocity of light in \textit{in vacuo} and $G$ is the Newton constant.

\section{Anisotropic stress tensor}
To start with, we write down a generalized form of the spherically-symmetric Painleve-Gullstrand (PG) line-element  \cite{HHR1, HHR2, IC}
		 \begin{equation}
  ds^{2} = -\left[1 - \frac{2m(t,r)}{r}\right]dt^{2} - 2\sqrt{\frac{2m(t,r)}{r}}dt dr + dr^{2} + r^{2} d \Omega^{2}, 
 \label{2.1}
 \end{equation}
 where $M(t,r)$ is the mass function and $d \Omega^{2}$ stands for the metric of the unit two-sphere. Compared to the standard PG spacetime (which is the Schwarzschild metric in different coordinates), we are dealing here with a dynamical geometry, where the mass function depends both on time and the radial coordinate. 

 The Einstein equations 
 		 \begin{equation}
		G_{ab} \equiv R_{ab} - \frac{1}{2}g_{ab}R = 8\pi T_{ab}
 \label{2.2}
 \end{equation}
give us the connection between the geometry (the Einstein tensor) and matter (the stress tensor $T_{ab}$). For the metric (2.1) one obtains the following non-zero mixed components of $G_{ab}$ 
  \begin{equation}
	\begin{split}
	G^{t}_{~t} = -\frac{2m'}{r^{2}},~~~ G^{r}_{~r} = -\frac{2m'}{r^{2}} - \frac{\sqrt{2}\dot{m}}{r\sqrt{rm}} ,~~~G^{r}_{~t} = \frac{2\dot{m}}{r^{2}},~~~G^{t}_{~r} = 0,\\~~~G^{\theta}_{~\theta} = G^{\phi}_{~\phi} = -\frac{{m''}}{r} - \frac{\sqrt{2}\dot{m'}}{2\sqrt{mr}} - \frac{\sqrt{2r}\dot{m}(m - rm')}{4mr^{2}\sqrt{m}}, 
\label{2.3}	
\end{split}
 \end{equation}
where $m' \equiv \partial m(t,r)/\partial r$ and $\dot{m} \equiv \partial M(t,r)/\partial t$. As the source of the geometry (2.1) we employ an energy-momentum tensor corresponding to an imperfect fluid \cite{HC, GRM}
	  \begin{equation}
	 T_{ab} = (p_{t} + \rho) u_{a} u_{b} + p_{t} g_{ab} + (p_{r} - p_{t}) n_{a}n_{b},
 \label{2.4}
 \end{equation}
where $\rho$ is the energy density of the fluid, $p_{r}$ is the radial pressure, $p_{t}$ are the transversal (tangential) pressures and $n^{a}$ is a unit spacelike vector orthogonal to $u^{a}$. We have $u_{a} n^{a} = 0,~u_{a} u^{a} = -1,~n_{a} n^{a} = 1$. In the geometry (2.1) one chooses a 4-velocity of the form (the fluid is expanding, so $u^{r}>0$)
	  \begin{equation}
		u^{a} = \left(1, \sqrt{\frac{2m(t,r)}{r}}, 0, 0\right), ~~~~u_{a} = (-1, 0, 0, 0)
 \label{2.5}
 \end{equation}
whence, keeping in mind the above relations, we get
	  \begin{equation}
		n^{a} = (0, 1, 0, 0),~~~n_{a} = \left(-\sqrt{\frac{2m(t,r)}{r}}, 1, 0, 0\right).
	\label{2.6}
	\end{equation}
		
\section{The mass function}
From now on we write symply $m$ instead of $m(t,r)$, for simplicity.
From Eqs.2.2, 2.3 and 2.4 we find that $G^{t}_{~r} = 0$ is an identity and 
\begin{equation}
-T^{t}_{~t} + T^{r}_{~r} = \rho + p_{r} = -\frac{\sqrt{2}\dot{m}}{8\pi r\sqrt{mr}}
 \label{3.1}
 \end{equation}
gives the same information as that obtained from the equation for $T^{r}_{~t}$, namely
\begin{equation}
 -(\rho + p_{r})\sqrt{\frac{2m}{r}} = \frac{\dot{m}}{4\pi r^{2}}
 \label{3.2}
 \end{equation}
It is worth noting that we have four unknown functions ($\rho, p_{r}, p_{t}, m$) and three equations 
\begin{equation}
\rho = \frac{m'}{4\pi r^{2}},~~~p_{r} = - \frac{m'}{4\pi r^{2}} - \frac{\sqrt{2}\dot{m}}{8\pi r\sqrt{mr}}
 \label{3.3}
 \end{equation}
and the last equation from (2.3), $G^{\theta}_{~\theta} = 8\pi p_{t}$. 

We plan to apply our model to high density stars (neutron stars, strange quark stars). Usually, an equation of state connecting the energy density and pressure is introduced for to solve the system of equations. We stick  to the Vertogradov, Ovgun and Shatov equation of state \cite{VOS} for the MIT bag model (see also \cite{VO, HC1, AM}), which is suitable for strange quark matter. However, our fluid is anisotropic ($p_{r} \neq p_{t}$) and, therefore, we propose a slightly modified form of it
\begin{equation}
p_{r} = \frac{1}{3}\left(\rho - 4b\right).
 \label{3.4}
 \end{equation}
We now insert the expressions of $\rho$ and $p_{r}$ in terms of $m$ and its derivatives from(3.3) in (3.4) and obtain the equation for the mass function
\begin{equation}
\dot{m}\sqrt{\frac{r}{2m}} + \frac{4m'}{3} - 2k^{2}r^{2} = 0,~~~~k^{2} \equiv 8\pi b/3.
 \label{3.5}
 \end{equation}
Let us seek a mass function of the form
\begin{equation}
m(t,r) = \frac{r^{3}}{2}g(t),
 \label{3.6}
 \end{equation}
where $g(t)$ is a positive function. When (3.6) is introduced in (3.5) one obtains
\begin{equation}
\frac{d}{dt}\sqrt{g(t)} + 2g(t) - 2k^{2} = 0,
 \label{3.7}
 \end{equation}
which gives us
\begin{equation}
g(t) = k^{2}tanh^{2}[2k(t - \alpha)],
 \label{3.8}
 \end{equation}
where $\alpha$ is a constant of integration, with $g(\alpha) = 0$. The mass function can be written as 
\begin{equation}
m(t,r) = \frac{k^{2}}{2}r^{3}tanh^{2}[2k(t - \alpha)].
 \label{3.9}
 \end{equation}
From now on we choose $\alpha = 0$, for simplicity. For $t\geq 0$ and $r = r_{0} = const.$, the mass function grows from zero to $k^{2}r_{0}^{3}/2 = (4\pi/3) r_{0}^{3}b$, $b \approx 10^{35}erg/cm^{3}$ being the bag constant. For $t>>1/2k \approx 10^{-4}s$ we have $tanh(2kt) \approx 1$ and $m$ is practically constant (at constant $r$). We found that $m(t,r_{0})$ is a monotonic function, having an inflexion point where $sinh(kt) = \sqrt{2}/2$. 

One notes that, using (3.9), the potential $g_{tt}$ from (2.1) can be written as
 \begin{equation}
-g_{tt} = 1 - \frac{r^{2}}{R^{2}}tanh^{2}(2kt),
 \label{3.10}
 \end{equation}
where $R = \sqrt{3/8\pi b}$. One observes that $-g_{tt}$ is non-negative. There is no an event horizon because
 \begin{equation}
g_{tt} - \frac{g^{2}_{tr}}{g_{rr}} = -1 \neq 0.
 \label{3.11}
 \end{equation}

\section{Energy density and pressures}
Let us find now the parameters of the anisotropic fluid once the mass function has been determined. One obtains from (3.9) that
\begin{equation}
8\pi \rho = 3k^{2}tanh^{2}(2kt),~~~~8\pi p_{r} = 8\pi p_{t} = -3k^{2} - \frac{k^{2}}{cosh^{2}(2kt)}.
 \label{4.1}
 \end{equation}
An important observation is that $\rho,~p_{r},~p_{t}$ depend only on time and not on the radial coordinate. They are constant on a hypersurface of constant time; that is not a consequence of the separation of variables in the expression of $m(t,r)$. Moreover, one observes that $p_{r} = p_{t}$, i.e. the fluid is isotropic, with equally radial and transversal pressures.  
It is clear that only $\rho$ is positive for any time. 

 It is worth noting that $\rho<|p_{r}|$ and the dominant energy condition (DEC) is not satisfied. The same is valid for the other energy conditions. However, that is valid for short times. If $t>>1/2k$, the time dependent term in the expressions of the pressures  may be neglected and we obtain $\rho = -p_{r} = -p_{t} = 3k^{2}$. The expanding isotropic fluid has a de Sitter type stress tensor, with $\Lambda = k^{2} = 8\pi b/3$ and the radius of curvature $R = 1/k \approx 30 Km$, a value of the order of the radius of a neutron star or a quark star. Noting that a de Sitter geometry has been also proposed in \cite{HC1} for the metric inside a hadron. Another interesting property of the energy density and pressures from (4.1) is that they are time-symmetric such that the model may be extended for $t<0$, when the system is supposed to collapse.

\section{Kinematical quantities and radial geodesic}	
From the velocity field (2.5) one finds that the corresponding covariant acceleration is vanishing
	 \begin{equation}
a^{b} = u^{a}\nabla_{a}u^{b} = 0.
 \label{5.1}
 \end{equation}
That means $u^{r} = \sqrt{2m(t,r)/r}$ is tangent to the outward radial geodesics, such that the observer is comoving with the fluid and the energy flux densities $q^{a}$ are vanishing. It can be shown that the scalar expansion $\Theta = \nabla_{b}u^{b}$ given by
	\begin{equation}
	\Theta = \frac{\sqrt{2}}{2}~\frac{3m(t,r) + rm'(t,r)}{r^{2}}\sqrt{\frac{r}{m(t,r)}}
 \label{5.2}
 \end{equation}
is positive and depends only on time, namely $\Theta = 3k~tanh(2kt)$.  Moreover, the shear tensor $\sigma^{a}_{~b} = 0$ for our geometry (2.1), because the mass function satisfies $3m - rm' = 0$. In addition, the scalar curvature is given by
	\begin{equation}
	R^{a}_{~a} = 12k^{2}, 
 \label{5.3}
 \end{equation}
  which is constant and $R^{a}_{~a}$ acquires exactly the de Sitter value $12k^{2}$.

We already noticed that our observer is expanding freely with the fluid, like the particles in the time dependent de Sitter spacetime from Cosmology. Therefore, the geodesics could be computed from (see 2.5)
\begin{equation}
\frac{dr}{dt} = \sqrt{\frac{2m(t,r)}{r}} = kr ~tanh(2kt), ~~~t>0.
 \label{5.4}
 \end{equation}
We have in (5.4) $dr/dt<1$ because $kr<1$ for $r\leq R<30 Km$.
After a simple integration, the trajectory can be written as
\begin{equation}
r(t) = r_{min}\sqrt{cosh(2kt)},
 \label{5.5}
 \end{equation}
with $r_{min} = r(0)$, a value which is aiming to be determined in a future work. From (5.4) one observes that $dr/dt$ is an odd function w.r.t. time and the formula is not valid for $t<0$, when the fluid should collapse. 

\section{Conclusion}
We propose a model for the gravitational expanding matter having an anisotropic fluid as the source (with an equation of state similar to that one from the MIT Bag Model applied to hadrons or strange quark stars). The observer is comoving with the isotropic fluid and feels no energy flux density. The energy density of the fluid is always positive but the pressures are negative. The scalar curvature corresponding to the geometry (2.1) with $m(t,r)$ from (3.6) is constant and positive, even though the geometry is time dependent.
When $t \rightarrow \infty$ (or $t>>1/2k$), the metric is exactly de Sitter, with $\Lambda = k^{2}$.

\end{document}